\documentclass[sigconf,screen]{acmart}

\usepackage{bm}
\usepackage{braket}
\usepackage{mathrsfs}
\usepackage[commandnameprefix=ifneeded,authormarkuptext=name,authormarkup=none,final]{changes}
\definecolor{red}{RGB}{255, 0, 0}
\definecolor{blue}{RGB}{0, 0, 255}
\definechangesauthor[name={Gianluca}, color={red}]{GL}
\definechangesauthor[name={Yorick}, color={blue}]{YLAS}
\DeclareMathAlphabet{\mathscrbf}{OMS}{mdugm}{b}{n}
\AtBeginDocument{%
  }

\setcopyright{acmlicensed}
\copyrightyear{2018}
\acmYear{2018}
\acmDOI{XXXXXXX.XXXXXXX}

\acmConference[PASC24]{Make sure to enter the correct
  conference title from your rights confirmation emai}{June 03--05,
  2024}{Zurich, Switzerland}
\acmISBN{978-1-4503-XXXX-X/18/06}

\newcommand*{\matr}[1]{\mathbf{#1}}
\newcommand*{\vect}[1]{\bm{#1}}

\begin{document}

\title[Saddle Point Search Algorithms for Excited Electronic States with Self-Interaction Correction]{Saddle Point Search Algorithms for Variational Density Functional Calculations of Excited Electronic States with Self-Interaction Correction}

\author{Yorick Leonard Adrian Schmerwitz}
\email{yla1@hi.is}
\orcid{0000-0001-6277-0359}
\author{N\'uria Urgell Oll\'e}
\author{Gianluca Levi}
\orcid{0000-0002-4542-0653}
\author{Hannes J\'onsson}
\orcid{0000-0001-8285-5421}
\affiliation{%
  \institution{Science Institute and Faculty of Physical Sciences, University of Iceland}
  \streetaddress{Dunhagi 5}
  \city{Reykjavík}
  \country{Iceland}
  \postcode{107}
}

\renewcommand{\shortauthors}{Schmerwitz \textit{et al.}}

\begin{abstract}
Excited electronic states of molecules and solids play a fundamental role in fields such as catalysis and electronics. In electronic structure calculations, excited states typically correspond to saddle points on the surface described by the variation of the energy as a function of the electronic degrees of freedom. A direct optimization algorithm based on generalized mode following is presented for density functional calculations of excited states. While conventional direct optimization methods based on quasi-Newton algorithms usually converge to the stationary point closest to the initial guess, even minima, the generalized mode following approach systematically targets a saddle point of a specific order $l$ by following the $l$ lowest eigenvectors of the electronic Hessian up in energy. This approach thereby recasts the challenging saddle point search as a minimization, enabling the use of efficient and robust minimization algorithms. The initial guess orbitals and the saddle point order of the target excited state solution are evaluated by performing an initial step of constrained optimization freezing the electronic degrees of freedom involved in the excitation. In the context of Kohn-Sham density functional calculations, typical approximations to the exchange-and-correlation functional suffer from a self-interaction error. The Perdew and Zunger self-interaction correction can alleviate this problem, but makes the energy variant to unitary transformations in the occupied orbital space, introducing a large amount of unphysical solutions that do not fully minimize the self-interaction error. An extension of the generalized mode following method is proposed that ensures convergence to the solution minimizing the self-interaction error.
\end{abstract}

\ccsdesc[500]{Applied computing~Chemistry}
\ccsdesc[500]{Applied computing~Physics}

\keywords{Excited states, variational calculations, density functional calculations, self-interaction correction, saddle point searches, generalized mode following}

\received{20 February 2007}
\received[revised]{12 March 2009}
\received[accepted]{5 June 2009}

\maketitle
\section{Introduction}
The interaction of light with electrons forms the basis of the field of photochemistry.\cite{Balzani2015} This interaction is of vital importance in many state-of-the-art technologies, such as solar cells, which convert light to electrical energy\cite{Rmaroli2016}, or in photocatalysts,\cite{Romero2017} which harness light to enable chemical reactions storing energy into chemical bonds. These advancements pave the way for sustainable human development. 

When electrons in molecules and solids interact with light, they are promoted from the ground electronic state to an excited state. The excited electronic state has a limited life time before the electronic system returns back to its ground state, which represents a state of equilibrium. Quantum chemistry calculations of molecules and solids typically use the Born-Oppenheimer approximation, which partitions the total wave function into a product of an electronic and a nuclear wave function. The electronic states are then evaluated by solving an approximate form of the electronic Schröd-inger equation for the electronic wave function, which is parametrically dependent on the nuclear positions. Among several approaches, Kohn-Sham\cite{Kohn1965} (KS) density functional theory\cite{Hohenberg1964} (DFT) has been one of the most successful, as it combines relatively low computational cost with reasonable accuracy for a large range of systems. DFT and KS functionals have originally been formulated for calculations of the ground state. A time-dependent extension (TDDFT) has been developed for excited states,\cite{Runge1984} but it is in many ways limited.\cite{Levine2006} Alternatively, the excited states of an electronic system can also be obtained as solutions to the time-independent electronic problem, in an approach analogous to ground state calculations. While the ground state is the global minimum on the surface defined by the variation of the energy as a function of the electronic degrees of freedom, excited states are represented by stationary points higher in energy than the ground state.\cite{Burton2022, Hait2021, Levi2020b, Perdew1985} Minimizing the energy within KS DFT calculations to find the ground electronic state is a well known problem. However, finding excited electronic state solutions involves the search for saddle points on the energy surface, which poses additional challenges. 

\added[id=YLAS, comment={R1C1}]{Ref. \cite{Schmerwitz2023} presents a direct orbital optimization method that can be used to find saddle points on the electronic energy surface for excited state calculations. Here, we present an extension of this approach that enables the inclusion of self-interaction correction (SIC) as proposed by Perdew and Zunger\cite{Perdew1981}, which alleviates the self-interaction error (SIE) inherent in practical, approximate density functionals.} All algorithms presented in this article have been implemented in the GPAW software and are publicly available.\cite{Mortensen2023}

The article is structured as follows. First, the KS formulation of DFT and algorithms for calculations of the ground electronic state are introduced. Then, Perdew-Zunger SIC is discussed. Afterward, the connection between stationary points on the KS electronic energy surface including SIC and excited states is illustrated using a model based on the H$_{2}^{-}$ ion described with a minimal basis set. Excited state saddle point search algorithms are presented and then extended to include SIC. The article ends with a conclusion section.
\section{Kohn-Sham density functional calculations}
In Hartree-Fock theory and KS DFT, the wave function of an electronic system is described as a single Slater determinant of orthonormal molecular orbitals. A stationary electronic state is obtained by finding a set of optimal molecular orbitals that makes the energy stationary. Most commonly, calculations on molecular systems employ the linear combination of atomic orbitals (LCAO) approach, where the molecular orbitals $\vect{\psi} = \left(\ket{\psi_1}, \ldots, \ket{\psi_{M}}\right)$ are expanded in terms of generally non-orthonormal localized basis functions $\vect{\phi} = \left(\ket{\phi_1} \ldots, \ket{\phi_{M}}\right)$\, according to
\begin{equation}
\vect{\psi} = \vect{\phi}\matr{C}\,,
\end{equation}
with $\matr{C}$ being an $M \times M$ matrix of expansion coefficients. The objective is then to find an optimal set of coefficients, which makes the total energy stationary.

In KS DFT, the energy is given by (in atomic units)
\begin{align}\label{eq:ks-energy}
    E_\mathrm{KS}\left[n\right] = T_{\mathrm{S}}\left[n\right] & + \int d\mathbf{r}v_{\mathrm{ext}}\left(\mathbf{r}\right)n\left(\mathbf{r}\right)\\
    & + \frac{1}{2}\int\int d\mathbf{r}d\mathbf{r}'\frac{n\left(\mathbf{r}\right)n\left(\mathbf{r}'\right)}{\left|\mathbf{r} - \mathbf{r}'\right|} + E_{\mathrm{XC}}\left[n\right]\,.\nonumber
\end{align}
$T_{\mathrm{S}}\left[n\right]$ and $n(\mathbf{r})$ are the kinetic energy of non-interacting electrons and the total electron density, respectively, which are given in terms of the molecular orbitals
\begin{align}\label{eq:ekin}
T_{\mathrm{S}}\left[n\right] = -\frac{1}{2} \sum_{i}^{M} f_i \bra{\psi_{i}}\nabla^{2}\ket{\psi_{i}}\,,
\end{align}
\begin{align}\label{eq:totdens}
n(\mathbf{r}) \sum_{i}^{M} f_i \left|\psi_{i}(\mathbf{r})\right|^{2}\,, 
\end{align}
where $f_i$ is the occupation number of orbital $i$, $\nabla^{2}$ involves a summation over second-order derivatives of the spatial electronic coordinates, and $\psi_{i}(\mathbf{r})=\braket{\mathbf{r}|\psi_{i}}$, with $\mathbf{r}$ being the electronic coordinates. $\bra{}\ket{}$ indicates integration over the electronic coordinates. In general,
\begin{equation}
    \bra{\psi_{i}\psi_{j}}\matr{\hat{O}}\ket{\psi_{j}\psi_{i}} = \int d\mathbf{x}d\mathbf{x}'\psi_{i}^{*}\left(\mathbf{x}\right)\psi_{j}^{*}\left(\mathbf{x}'\right)\matr{\hat{O}}\psi_{j}\left(\mathbf{x}\right)\psi_{i}\left(\mathbf{x}'\right)\,,
\end{equation}
with $\matr{\hat{O}}$ being an arbitrary operator, $\mathbf{x}$ a convolution of the spatial and spin electronic coordinates, and $\psi(\mathbf{x})$ is a product of $\psi(\mathbf{r})$ and one of two orthonormal spin functions, $\alpha$ and $\beta$. The second term in eq.~\ref{eq:ks-energy} is the external potential energy corresponding to the Coulomb electron-nuclear attraction due to the nuclear external potential, $v_{\mathrm{ext}}$. The third term is the Coulomb electron-electron repulsion energy, and the fourth term is the energy due to electron-electron exchange and Coulomb correlation, which in practical calculations is described by an approximate functional of the electron density. 

\section{Perdew-Zunger self-interaction correction}
\added[id=YLAS, comment={R1C4}]{If the exchange-and-correlation functional in eq.~\ref{eq:ks-energy} is replaced by Fock exchange, the Hartree-Fock expression of the energy is recovered}
\begin{align}\label{eq:HF}
    E_{\mathrm{HF}} & = \sum_{i}^{N_{e}}\left(\bra{\psi_{i}}-\frac{1}{2}\nabla^{2}\ket{\psi_{i}} - \bra{\psi_{i}}\sum_{\alpha}^{N_A}\frac{Z_{\alpha}}{\left|\mathbf{r} - \mathbf{r_{\alpha}}\right|}\ket{\psi_{i}}\right)\\
    & + \frac{1}{2}\sum_{ij}^{N_{e}}\left(\bra{\psi_{i}\psi_{j}}\frac{1}{\left|\mathbf{r} - \mathbf{r}'\right|}\ket{\psi_{i}\psi_{j}} - \bra{\psi_{i}\psi_{j}}\frac{1}{\left|\mathbf{r} - \mathbf{r}'\right|}\ket{\psi_{j}\psi_{i}}\right)\,,\nonumber
\end{align}
where $N_{e}$ is the number of electrons, $N_{A}$ the number of nuclei, and $Z_{\alpha}$ the charge of nucleus $\alpha$. The first and second one-electron terms in eq.~\ref{eq:HF} correspond to the kinetic energy and the electron-nuclear attraction, respectively, while the first and second two-electron terms are the Coulomb and exchange interaction, respectively. The sum over the two-electron terms explicitly includes those integrals where the electronic indices $i$ and $j$ are equal. These Coulomb and exchange self-interaction terms are identical and therefore, cancel exactly.

While the KS formalism very efficiently recovers some of the Coulomb correlation missing in the Hartree-Fock approach, it describes exchange in a semi-local way, lifting the balance between the exchange and non-local Coulomb self-interaction terms, introducing the infamous self-interaction error. The self-interaction error is the root cause of several errors in KS calculation, such as excessive delocalization of charge observed for instance in anions.\cite{zhang_2016} This spurious self-interaction is even present in single-electron systems and makes the long-range form of the potential acting on the electron decay too fast, deviating from the correct $\frac{1}{r}$ behavior. This incorrect decay introduces large errors in, e.g., calculations of Rydberg excited states.\cite{Sigurdarson2023}

Self-interaction correction (SIC) is a non-trivial pursuit since self-interaction is inherently a many-body effect. Perdew and Zunger have proposed a single-electron SIC\cite{Perdew1981} by subtracting the difference of the self-interaction terms in the sum of Coulomb integrals and the exchange-and-correlation functional
\begin{align}\label{eq:sic}
    E_{\mathrm{KS}}^{\mathrm{SIC}}&\left[n_{1}, n_{2} \dots n_{N_{e}}\right] = E_{\mathrm{KS}}\left[n\right] \nonumber \\ 
    & - \frac{1}{2}\sum_{i}^{N_{e}}\left(\int d\mathbf{r}d\mathbf{r}'\frac{n_{i}\left(\mathbf{r}\right)n_{i}\left(\mathbf{r}'\right)}{\left|\mathbf{r} - \mathbf{r}'\right|} + E_{\mathrm{XC}}\left[n_{i}\right]\right)\,,
\end{align}
where $n_{i} = \left|\psi_{i}\right|^{2}$ is the density of orbital $i$ and $n = \sum_{i}^{N_{e}}n_{i}$. Eq.~\ref{eq:ks-energy} expresses the energy as a functional of the total electron density $n$ given by eq. \ref{eq:totdens}, which is therefore invariant to a unitary transformation of the occupied orbitals. As eq.~\ref{eq:sic} introduces terms that depend on the individual orbital densities, the energy is not invariant to unitary transformation anymore if Perdew-Zunger SIC is applied.
\section{Ground electronic state}
Finding the ground electronic state involves first choosing an LCAO basis set, usually Gaussian functions taken from a published set whose coefficients have been fitted to atomic orbitals. The incomplete nature of this set introduces an error that can be controlled by choosing an appropriate number and quality of functions. This definition is followed by iterative minimization of eq.~\ref{eq:sic}, effectively yielding a unitary transformation matrix $\matr{U}$ 
\begin{equation}
    \label{eq:unitary-transformation}
    \matr{C}_{\mathrm{min}} = \matr{C}_{0}\matr{U}
\end{equation}
transforming the coefficient matrix of the initial orbitals, $\matr{C}_{0}$\,, into that of the orbitals, $\matr{C}_{\mathrm{min}}$\,, which minimize the KS energy. Any minimization algorithm can be used for this purpose, but the most commonly employed is the self-consistent field (SCF) method,\cite{Lehtola2020} which iteratively forms and diagonalizes the KS Hamiltonian matrix with elements
\begin{equation}
    H_{ij} = \bra{\psi_{i}}\matr{\hat{H}}_{\mathrm{KS}}\ket{\psi_{j}}\,,
\end{equation}
where $\mathbf{\hat{H}}_{\mathrm{KS}}$ is the Kohn-Sham Hamiltonian
\begin{equation}\label{eq:ks-ham}
    \mathbf{\hat{H}}_{\mathrm{KS}} = -\frac{1}{2}\nabla^{2} + v_{\mathrm{ext}}(\mathbf{r}) + \int d\mathbf{r'}\frac{n(\mathbf{r'})}{\left|\mathbf{r} - \mathbf{r'}\right|} + v_{\mathrm{XC}}(\mathbf{r})\,.
\end{equation}
Imposing orthonormality between the molecular orbitals leads to
\begin{equation}
    \mathbf{\hat{H}}_{\mathrm{KS}}\ket{\psi_i} = \sum_{j}\epsilon_{ij}\ket{\psi_{j}}\,,
\end{equation}
where the $\epsilon_{ij}$ are Lagrange multipliers. Each diagonalization of the KS Hamiltonian alters the orbitals, in turn altering the KS Hamiltonian. This process is repeated until the two steps have sufficiently small impact on each other, which is when self-consistency is reached. At the SCF solution, the matrix of Lagrange multipliers is diagonal, and the diagonal elements represent the energy of canonical one-particle states (the KS canonical orbitals).

Despite many improvements to this simple SCF algorithm over the decades, a more direct approach which directly finds the unitary transformation matrix $\matr{U}$ in eq.~\ref{eq:unitary-transformation} is typically found to outperform SCF both in stability and efficiency.\cite{Neese2000, Voorhis2002} This direction optimization (DO) method typically uses an exponential transformation\cite{Hutter1994, VandeVondele2003}
\begin{equation}
    \matr{U} = e^{\boldsymbol{\kappa}}, \mathrm{\ subject\ to\ } \boldsymbol{\kappa} = -\boldsymbol{\kappa}^{\dag}
\end{equation}
to parameterize $\matr{U}$. To comply with the orbital orthonormality constraints, the exponentiated matrix $\boldsymbol{\kappa}$ is required to be anti-Hermitian. The matrix $\boldsymbol{\kappa}$ contains pairwise orbital rotations mixing the occupied (o) orbitals with the occupied (oo) and virtual (ov) orbitals and the virtual (v) orbitals with the virtual (vv) orbitals
\begin{equation}
    \boldsymbol{\kappa} = \begin{pmatrix}\boldsymbol{\kappa}_{\mathrm{oo}} & \boldsymbol{\kappa}_{\mathrm{ov}}\\ -\boldsymbol{\kappa}_{\mathrm{ov}}^{\dag} & \boldsymbol{\kappa}_{\mathrm{vv}}\end{pmatrix}\,.
\end{equation}
Minimizing the energy with respect to the elements of $\matr{U}$ is equivalent to solving the SCF equations. While the unitary matrices form a non-linear space, the space of anti-Hermitian matrices is linear. Since any stationary point in the linear space is also stationary in the corresponding exponential space, linear optimization techniques can be applied using the orbital rotations in $\boldsymbol{\kappa}$ as degrees of freedom. To do so, the gradient of the KS energy with respect to the elements of $\boldsymbol{\kappa}$ needs to be known
\begin{align}
    \label{eq:gradient}
    \frac{\partial E}{\partial \kappa_{ij}} & = \frac{2 - \delta_{ij}}{2}\left[\int_{0}^{1}dte^{t\boldsymbol{\kappa}}\matr{L}e^{-t\boldsymbol{\kappa}}\right]_{ji}\\
    & = \frac{2 - \delta_{ij}}{2}\left[\matr{L} + \frac{1}{2!}\left[\boldsymbol{\kappa}, \matr{L}\right] + \frac{1}{3!}\left[\boldsymbol{\kappa}, \left[\boldsymbol{\kappa}, \matr{L}\right]\right] + ...\right]_{ji}\nonumber\,.
\end{align}
Here, $\delta_{ij}$ is an element of a unit matrix, $\left[A, B\right] = AB - BA$ indicates the commutator of $A$ and $B$, and
\begin{equation}
    L_{ij} = \left(f_{i} - f_{j}\right)H_{ij} - f_{i}V_{ij} + f_{j}V_{ji}^{*}\,,
\end{equation}
where the $V_{ij}$ are the elements of the SIC potential matrix $\mathbf{V}$ 
\begin{equation}
    V_{ij} = \bra{\psi_{i}\left(\mathbf{r}\right)}\left(\int d\mathbf{r}'\frac{n_{i}\left(\mathbf{r}'\right)}{\left|\mathbf{r} - \mathbf{r}'\right|} + v_{\mathrm{XC}}\left[n_{i}\left(\mathbf{r}\right)\right]\right)\ket{\psi_{j}\left(\mathbf{r}\right)}\,.
\end{equation}
While the contribution to the gradient due to the Hamiltonian matrix always leads (at points on the energy surface different from stationary points) to a finite gradient in the ov subspace of $\boldsymbol{\kappa}$, the contribution due to the SIC potential introduces a finite gradient in the oo subspace of $\boldsymbol{\kappa}$ as well, which is not present if SIC is not used ($\matr{V}$ = 0). \added[id=YLAS, comment={R1C6}]{It is well known that this variance of the energy to unitary transformations in the oo subspace of $\boldsymbol{\kappa}$ due to SIC introduces unphysical local minima on the energy surface.\cite{Lehtola2016} A unitary-invariant formulation of SIC\cite{Trepte2021} is possible by using Fermi-L\"owdin orbitals,\cite{Lowdin1950} but significantly complicates the calculations.}

The right-hand side of eq.~\ref{eq:gradient} is a special case of the Baker-Campbell-Hausdorff formula\cite{Baker1905, Campbell1896, Campbell1897, Hausdorff1906}, which applies since the anti-Hermitian matrices form the Lie-algebra corresponding to the Lie-group of unitary matrices. This commutator expansion may be truncated at the first term ($\frac{\partial E}{\partial \kappa_{ij}} \approx \frac{2 - \delta_{ij}}{2}L_{ji}$) provided that the norm of $\boldsymbol{\kappa}$ is kept small (${\| \boldsymbol{\kappa} \|}  \ll 1$), which can be achieved by applying eq.~\ref{eq:unitary-transformation} in regular optimization step intervals to update the orbitals and set $\boldsymbol{\kappa}$ to zero. The following diagonal approximation of the electronic Hessian is available as a preconditioner for the chosen linear optimization method
\begin{equation}
    \label{eq:diagonal_hessian}
    \frac{\partial^{2}E}{\partial \kappa_{ij}^{2}} \approx 2\left(f_{j} - f_{i}\right)\left(\epsilon_{i} - \epsilon_{j}\right)\,,
\end{equation}
$\epsilon_{i}$ being the eigenvalue of the Hamiltonian matrix corresponding to orbital $i$. Any linear minimization method can be used, such as the efficient L-BFGS\cite{Nocedal1980} quasi-Newton method, which is based on a positive-definite model Hessian, with line search techniques.
\section{Stationary points of the self-interaction corrected Kohn-Sham energy surface}\label{sec:stat}
\begin{figure*}[h!]
    \centering
    \includegraphics[width = 0.65\textwidth]{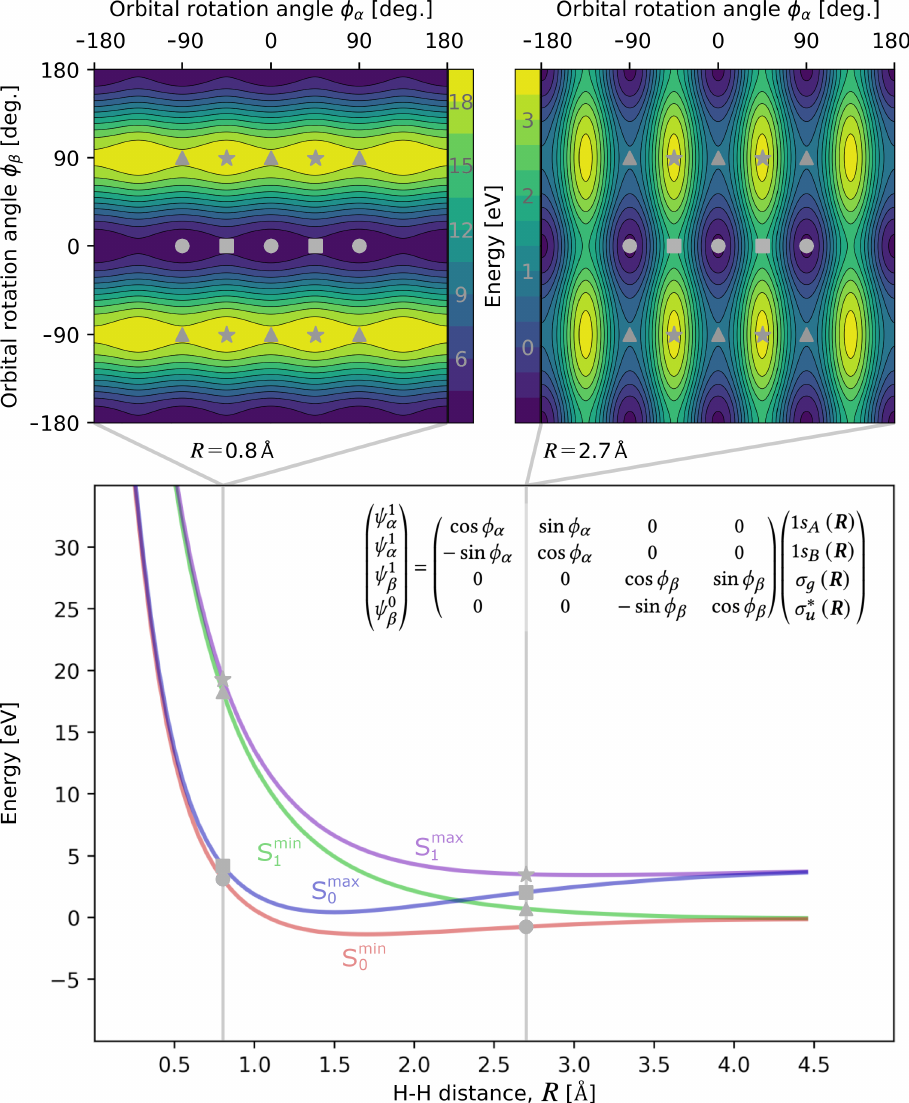}
    \caption{Energy as a function of bond length, $\vect{R}$, for the two states that can be obtained for the minimal-basis H$_{2}^{-}$ ion where two electrons are in the $\vect{\alpha}$ and one electron is in the $\vect{\beta}$ spin channel. The orbitals are related by rotation angles $\vect{\phi_{\alpha}}$, mixing two occupied orbitals, and $\vect{\phi_{\beta}}$ mixing an occupied and a virtual orbital, with respect to the orbitals of the ground state solution, S$\vect{_{0}^{\mathrm{min}}}$, corresponding to the localized orbitals $\vect{1s_{A}}$ and $\vect{1s_{B}}$ at H\textsubscript{A} and H\textsubscript{B}, respectively, in the $\vect{\alpha}$ channel and the delocalized bonding and antibonding orbitals $\vect{\sigma_{g}}$ and $\vect{\sigma_{u}^{*}}$, respectively, in the $\vect{\beta}$ channel. The S$\vect{_{1}}$ state is obtained by a single excitation in the $\vect{\beta}$ spin channel with respect to S$\vect{_{0}}$. The superscripts $\vect{0}$ and $\vect{1}$ indicate the orbital occupation, while the superscripts min and max indicate whether a minimum or maximum of the self-interaction error has been found, respectively.
    The left contour graph corresponds to a bond length of $\vect{R} \vect{= 0.8}$\,\AA. There, S$\vect{_{0}^{\mathrm{min}}}$ corresponds to a minimum (circles), while S$\vect{_{0}^{\mathrm{max}}}$ (squares) and S$\vect{_{1}^{\mathrm{min}}}$ (triangles) correspond to 1\textsuperscript{st}-order saddle points with the direction of negative curvature along $\vect{\phi_{\alpha}}$ and $\vect{\phi_{\beta}}$, respectively. S$\vect{_{1}^{\mathrm{max}}}$is represented by a 2\textsuperscript{nd}-order saddle point. The curvature of the surface along $\vect{\phi_{\alpha}}$ at S$\vect{_{0}^{\mathrm{min}}}$ is much smaller than along $\vect{\phi_{\beta}}$. The right contour graph corresponds to a stretched bond length of $\vect{R} \vect{= 2.7}$\,\AA. The locations of the stationary points persist, but the curvature changes significantly with the curvature along $\vect{\phi_{\alpha}}$ now being much larger than along $\vect{\phi_{\beta}}$.}
    \Description{The figure shows energy curves of four states labeled S$_{0}^{\mathrm{min}}$ (red), S$_{0}^{\mathrm{max}}$ (blue), S$_{1}^{\mathrm{min}}$ (green), and S$_{1}^{\mathrm{max}}$ (purple). The x-axis is labeled "H-H distance, R [\AA]". The y-axis is labeled "Energy [eV]". S$_{0}^{\mathrm{min}}$ shows a minimum at ca. 1.5\,\AA, S$_{0}^{\mathrm{max}}$ at ca. 1.7\,\AA, while the other two curves are monotonically decreasing. The two curves with subscripts "0" and "1" approach each other in the direction of decreasing bond distance, while the curves with superscripts "min" and "max" approach each other in the opposite direction. S$_{0}^{\mathrm{max}}$ and S$_{1}^{\mathrm{min}}$ cross between these limits. There are grey vertical lines at bond distances of 0.8 and 2.7\,\AA linked to contour graphs above the figure containing the curves. These graphs show two-dimensional energy surfaces. The x-label reads "Orbital rotation angle $\phi_{\alpha} [deg.]$", the y-label "Orbital rotation angle $\phi_{\beta} [deg.]$". Color bars for the contour levels are given between the two graphs. They are labeled "Energy [eV]". Several stationary points on these surfaces are labeled by symbols. A minimum in the center of both graphs at the origin is labeled with circles. Three rows of stationary points are indicated at values of $\phi_{\beta}$ of -90, 0 and 90 and values of $\phi_{\alpha}$ of -90, -45, 0, 45, and 90. From the top left (-90, 90) to the bottom right (90, -90), the labels correspond to\\
    triangle, star, triangle, star, triangle,\\
    circle, square, circle, square, circle,\\
    triangle, star, triangle, star, triangle.\\
    The curvatures of the graphs are visibly different, leading to ovaloid shape of the contours at the minima and maxima with the longer direction of the ovaloids oriented along the x-axis in the left graph at a bond distance of 0.8\,\AA, while it is oriented along the y-axis in the right graph at a distance of 2.7\,\AA. $\phi_{\alpha}$ and $\phi_{\beta}$ are defined at the top right of the figure containing the curves. The equation is the same as in the main text.}
    \label{fig:h2-pes}
\end{figure*}
\added[id=YLAS, comment={R1C2}]{While the methods presented in this article have been applied to larger systems such as Rydberg excited states of small molecules\cite{Sigurdarson2023} and the torsional energy curves of the ground and several excited states of ethylene,\cite{Schmerwitz2022} we focus here on the  H\textsubscript{2} molecule as an illustrative example system.} Figure~\ref{fig:h2-pes} shows the energy as a function of the distance between the hydrogen atoms, $R$, in the H$_{2}^{-}$ ion. The energy is obtained with the PBE-SIC\cite{Perdew1996, Perdew1997} density functional, a commonly used functional with SIC. A minimal basis set is employed meaning that two orbitals are present in each of the spin channels, $\alpha$ and $\beta$. The H$_{2}^{-}$ ion has three electrons, i.e.~the $\alpha$ spin channel is fully occupied, while the $\beta$ spin channel has one occupied and one virtual orbital. There are two electronic states that can be described with a minimal basis set. The ground state is the global minimum of the electronic energy and is weakly binding with a minimum at $R = 1.7$\,\AA. The second state is an excited state accessible from the ground state minimum by switching the occupation numbers in the $\beta$ spin channel with a monotonically decreasing energy curve. These two solutions become degenerate in the dissociation limit ($R \to \infty$). With the PBE-SIC functional, there are exactly two electronic degrees of freedom defined as
\begin{equation*}
    \begin{pmatrix}
        \vspace{1pt}
        \psi^{1}_{\alpha}\\
        \vspace{1pt}
        \psi^{1}_{\alpha}\\
        \vspace{1pt}
        \psi^{1}_{\beta}\\
        \vspace{1pt}
        \psi^{0}_{\beta}
    \end{pmatrix}
    =
    \begin{pmatrix}
            \cos{\phi_{\alpha}} & \sin{\phi_{\alpha}} & 0 & 0\\
            -\sin{\phi_{\alpha}} & \cos{\phi_{\alpha}} & 0 & 0\\
            0 & 0 & \cos{\phi_{\beta}} & \sin{\phi_{\beta}}\\
            0 & 0 & -\sin{\phi_{\beta}} & \cos{\phi_{\beta}} 
    \end{pmatrix}
    \begin{pmatrix}
        \vspace{1pt}
        1s_{A}\left(R\right)\\
        \vspace{1pt}
        1s_{B}\left(R\right)\\
        \vspace{1pt}
        \sigma_{g}\left(R\right)\\
        \vspace{1pt}
        \sigma^{*}_{u}\left(R\right)
    \end{pmatrix}\,,
\end{equation*}
where the superscripts 1 and 0 correspond to occupied and virtual orbitals, respectively, and the subscripts $\alpha$ and $\beta$ correspond to the two spin channels. $\phi_{\alpha}$ mixes two occupied orbitals and changes the energy because SIC is used, while $\phi_{\alpha}$ mixes an occupied and a virtual orbital and would change the energy even for the uncorrected functional. 

Figure~\ref{fig:h2-pes} also illustrates the electronic energy surface as a function of $\phi_{\alpha}$ and $\phi_{\beta}$ at $R = 0.8$\,\AA\ and $R = 2.7$\,\AA. The zero values of the rotations are chosen to reproduce the ground state orbitals, which are different for the two spin channels. In the $\alpha$ channel, SIC induces maximum localization of the orbitals yielding the $1s$ orbitals of the two hydrogen atoms $1s_{A}$ and $1s_{B}$. In the $\beta$ channel, the orbitals are mixed producing the bonding and anti-bonding orbitals $\sigma_{g} = \frac{1}{\sqrt{2}}\left(1s_{A} + 1s_{B}\right)$ and $\sigma_{u}^{*} = \frac{1}{\sqrt{2}}\left(1s_{A} - 1s_{B}\right)$, respectively. The locations of the stationary points of the surfaces are identical for both distances. A $\pm 45^{\circ}$ rotation converts between the localized and delocalized orbital sets. At $\phi_{\alpha} = \phi_{\beta} = 0^{\circ}$, the global minimum is observed, which minimizes all degrees of freedom including maximizing the SIC, which minimizes the SIE. This solution is hence called S$_{0}^{\mathrm{min}}$. Applying a $\pm 90^{\circ}$ rotation in either spin channel corresponds to swapping the orbitals in that spin channel. If a $\pm 90^{\circ}$ rotation is applied in $\phi_{\beta}$, S$_{1}^{\mathrm{min}}$ is reached, which is represented by a 1\textsuperscript{st}-order saddle point. This solution still minimizes the SIE. The same rotation in $\phi_{\alpha}$ has no effect since the two orbitals that are swapped are both occupied. Performing a $\pm 45^{\circ}$ rotation in $\phi_{\beta}$ does not lead to a stationary point, while doing so in $\phi_{\alpha}$ converts the localized occupied orbitals to delocalized orbitals, which maximizes the SIE. Hence, the corresponding solutions are termed S$_{0}^{\mathrm{max}}$ and S$_{1}^{\mathrm{max}}$, which are a first and a second order saddle points, respectively. The two pairs, S$_{0}^{\mathrm{min}}$/S$_{0}^{\mathrm{max}}$ and S$_{1}^{\mathrm{min}}$/S$_{1}^{\mathrm{max}}$, describe the same states and therefore, show qualitatively the same energy curves, the main quantitative differences being the location of the minimum of S$_{0}^{\mathrm{max}}$ at $R = 1.5$\,\AA, slightly shorter than for S$_{0}^{\mathrm{min}}$, and the energy in the dissociation limit, which is larger for S$_{0}^{\mathrm{max}}$ and S$_{1}^{\mathrm{max}}$ than for S$_{0}^{\mathrm{min}}$ and S$_{1}^{\mathrm{min}}$ due to the difference in the SIEs. In the $R \to 0$ limit, the energy becomes infinite, and the SIE vanishes since there is no difference between the localized and delocalized orbital sets, so the two pairs of solutions become identical. When comparing the electronic energy surfaces at short and long bond distances, it is evident that the curvatures change qualitatively. While the direction of lowest curvature at the ground state minimum points along $\phi_{\alpha}$ at short bond distance, it changes to point along $\phi_{\beta}$ at longer bond distances.

An error, the SIE should of course be minimized. The possibility of accidentally maximizing the SIE introduces several unphysical stationary points in the subspace of oo rotations of $\boldsymbol{\kappa}$, which have to be avoided in variational density functional calculations. It is advisable to start these calculations from initial orbitals which have been localized, for instance by calculating Wannier orbitals.\cite{Pipek1989} This problem is particularly challenging for excited state optimizations, as section~\ref{sec:ex-sic} shows.
%
\section{Excited electronic states in the absence of self-interaction correction}\label{sec:ex-no-sic}
In this section, calculations of excited electronic states are discussed without the use of SIC first. SIC is then considered in the next section. The ground electronic state is always represented by the global minimum on the KS energy surface, with or without SIC.

As illustrated in the upper panel of fig.~\ref{fig:h2-pes} for the H$_{2}^{-}$ ion, excited electronic states are represented by stationary points with higher energy than the global minimum, meaning they can correspond to saddle points. As such, a simple minimization is insufficient since there are directions in which the energy needs to be maximized. Prior to an excited state calculation, the ground state is evaluated. In the ground state, the orbitals are occupied in such a way that the energy is minimal according to the aufbau principle. An initial guess for an excited state optimization is obtained by choosing non-aufbau occupation numbers, for instance by promoting one electron from the highest occupied one-particle state to the lowest virtual state.

In principle, the SCF method is capable of converging to saddle points, but it has been designed for minimization and thus, the basin of attraction of saddle points is small for SCF, leading to this method minimizing along the directions where the energy should be maximized if the initial guess is not close to the target saddle point. The result is convergence to a lower-energy stationary point which can even be the ground state minimum. This process is commonly called variational collapse. There are two ways to approach this challenge. A method which counteracts variational collapse can be used together with the conventional SCF method or an optimization method can be employed that identifies the directions along which the energy needs to be maximized. The former approach works around variational collapse, while the latter method applies a systematic saddle point search.

The most commonly used method to counteract variational collapse is the maximum overlap method\cite{Gilbert2008, Barca2018} (MOM) which occupies those orbitals at each optimization step that overlap most with a reference orbital set, typically the occupied orbitals of either the initial guess or the previous optimization step. If the character of the occupied orbitals changes too much, the method changes the occupation numbers in an attempt to move back toward the target saddle point into its basin of attraction. This method has been shown not to work in the case of ov mixing of orbitals by ca. $45^{\circ}$ causing the collapse.\cite{Schmerwitz2023, Selenius2023}

Saddle point search methods are much less explored than minimization techniques. Quasi-Newton methods that allow for an indefinite model Hessian, such as the symmetric-rank 1,\cite{Murtagh1970} Powell,\cite{Powell1973} and Bofill\cite{Bofill1994} methods, and their limited-memory versions, have been used together with MOM (referred to as DO-MOM).\cite{Ivanov2023, Schmerwitz2022, Levi2020a, Levi2020b, Ivanov2021a, Ivanov2021b} Line search algorithms cannot be applied. In principle, MOM is obsolete when proper saddle point search algorithms are used, but it is typically still employed as a fail-safe due to its negligible computational cost. It is important to use the diagonal Hessian approximation in eq.~\ref{eq:diagonal_hessian} to provide an indefinite initial Hessian suitable for saddle point searches.

DO-MOM systematically converges to saddle points, but the nature of the stationary point is highly dependent on the initial guess for the saddle point search. The saddle point order, the number of directions of negative curvature at a saddle point, has emerged as an important characteristic of saddle points. While DO-MOM can converge to a saddle point of arbitrary order, a generalized mode following\cite{Schmerwitz2023} (GMF) method attempts to narrow down the target saddle point order first and is then guaranteed to converge to a saddle point of that order. The DO-GMF method inverts the projection of the total gradient, $\vect{g}$, onto the subspace formed by the eigenvectors, $\vect{v}$, of the electronic Hessian corresponding to its $l$ lowest eigenvalues, $\lambda$, according to
\begin{equation}
    \vect{g}^{\mathrm{\,mod}} = \displaystyle\ \vect{g} - 2\sum_{i = 1}^{l}\vect{v}_{i}\vect{v}_{i}^{\mathrm{T}}\vect{g}\,.
    \label{eq:gmf}
\end{equation}
The so-obtained modified gradient corresponds to an unknown modified objective function that shows a minimum where the energy shows the target saddle point, effectively recasting the challenging saddle point search as a minimization. As such, efficient minimization methods can be used for saddle point searches in this context, meaning that the Hessian update utilized for the quasi-Newton method needs not be able to develop an indefinite Hessian, so L-BFGS can be employed even for the saddle point search. By recasting the saddle point search problem into a minimization, the possibility of variational collapse is eliminated, thus MOM is neither needed nor beneficial. The lowest $l$ eigenpairs of the electronic Hessian are calculated with a generalized Davidson method using a finite difference method described in the appendix. This process introduces a higher computational cost than DO-MOM. The advantages of this method over DO-MOM are two-fold. It converges more robustly in cases of challenging excited states, as has been demonstrated previously,\cite{Schmerwitz2023} and it targets a specific saddle point order. The latter point is particularly useful in scenarios where multiple solutions on the KS energy surface emerge from one for a change of the molecular geometry\cite{Schmerwitz2023}.
\begin{figure*}[h!]
    \centering
    \includegraphics[width = \textwidth]{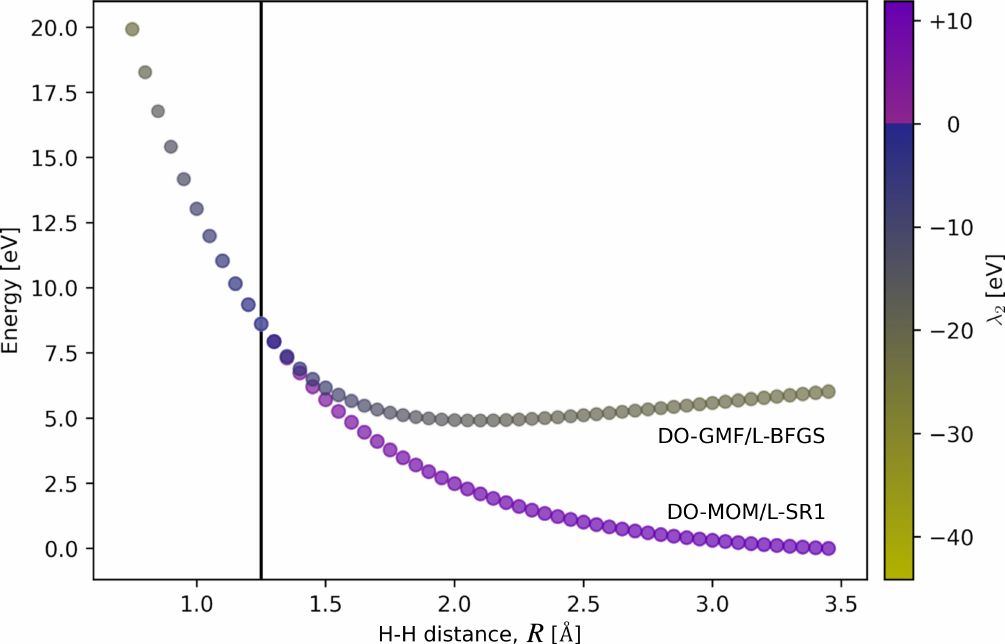}
    \caption{Energy of the lowest doubly excited state of H\textsubscript{2} as a function of the bond distance, $R$, calculated with DO-GMF and DO-MOM using sequential point acquisition (circles)\added[id=YLAS]{, the PBE functional and the aug-cc-pVDZ basis set}. A double excitation from the ground state is performed to initialize the excited state calculation at the first geometry with the smallest bond distance. The DO-GMF calculation targets a 2\textsuperscript{nd}-order saddle point. The points on the curves are colored according to the value of the second eigenvalue of the electronic Hessian, $\vect{\lambda_{2}}$, while the black vertical line marks where symmetry-broken solutions appear. Before that, both DO-MOM and DO-GMF converge on the 2\textsuperscript{nd}-order saddle point corresponding to a solution that conserves the symmetry of the Hamiltonian. After that, DO-MOM converges on a 1\textsuperscript{st}-order saddle point corresponding to the symmetry-pure solution giving an incorrect potential energy curve. Instead, the DO-GMF calculations keep converging on a 2\textsuperscript{nd}-order saddle point corresponding to a symmetry-broken solution with ionic character (H$^{+}$H$^{-}$/H$^-$H$^+$), thereby providing a more accurate potential energy curve.}
    \Description{An energy curve is shown in a discrete representation without connections between the points. The x-axis is labeled "H-H distance, $R$ [\AA]", the y-axis "Energy [eV]". The individual points are colored. A color bar is given on the right of the figure containing the curve labeled "$\lambda_{s}$ [eV]". The colors range from yellow to blue for negative values. There is a jump in colors from blue to purple at 0 clearly indicating the separation between negative and positive signs. The positive-sign colors range from a lighter to a darker shade of purple A black vertical line at about 1.25\,\AA, which is at about one fifth of the x-axis marks the point at which two curves emerge from one. The points gradually change color from yellow to blue before the vertical line. After the vertical line, the lower curve abruptly changes to purple, while the upper curve remains blue. The color of the lower curve proceeds to darken slightly, while the upper curve's color changes back toward yellow.}
    \label{fig:h2-sp-orders}
\end{figure*}

To speed up sequential excited state calculations at different molecular geometries, the initial guess for the saddle point search is not always restarted from the ground state orbitals, which may not even be available if the ground state is not of interest in the study. Instead, the occupied orbitals of the converged excited state solution at the previous geometry are used. This strategy is referred to as a sequential point acquisition algorithm. 

Figure~\ref{fig:h2-sp-orders} shows a case in an excited state energy curve of the H\textsubscript{2} molecule, obtained with the PBE density functional and \added[id=YLAS]{the aug-cc-pVDZ} basis set,\cite{Dunning1989, Kendall1992, Woon1994} where multiple excited state solutions emerge from one as the curve is evaluated starting from the smallest distance between the hydrogen atoms. The initial guess at the smallest distance is created by two simultaneous excitations of electrons from the highest occupied single-particle state to the lowest virtual one in each spin channel. Subsequent points use a sequential point acquisition algorithm. The energy curves are colored according to the second-lowest eigenvalue of the electronic Hessian, $\lambda_{2}$, which can be evaluated with the generalized Davidson method. It is sufficient to monitor this eigenvalue since the lower eigenvalue is always negative for these solutions and the higher eigenvalues are always positive. There is a point on these curves indicated by the black vertical line at which two solutions emerge from one, sometimes referred to as a catastrophe. For distances between the two hydrogen atoms below this critical distance, only one solution exists. This solution has negative $\lambda_{2}$, meaning that it corresponds to a 2\textsuperscript{nd}-order saddle point on the electronic energy surface. At the critical point, $\lambda_{2}$ becomes zero. After the critical point, $\lambda_{2}$ can either become positive, changing the saddle point order and leading to a set of 1\textsuperscript{st}-order saddle points, or $\lambda_{2}$ can become negative again, conserving the saddle point order and leading to a set of 2\textsuperscript{nd}-order saddle points with higher energy. The lower-energy solutions conserve the symmetry of the KS Hamiltonian, resulting in delocalization of the electron density over the molecule. This artificial delocalization stabilizes the system too much, yielding a too low energy. The higher-energy solutions, on the other hand, break the symmetry of the electron density with respect to the inversion center of the molecule. As such, the electron density becomes localized at one of the two hydrogen atoms. Due to the resulting electrostatic interaction between the positively and negatively charged hydrogen ions, this curve shows a minimum at a finite distance, while the symmetry-conserving energy curve decreases monotonically. It is known from high-level quantum chemistry calculations that the energy curve corresponding to the symmetry-broken solution is the qualitatively correct one.

Using DO-MOM (or SCF-MOM), the calculations systematically converge to the symmetry-conserving energy curve past the critical point. To break the symmetry and obtain the correct energy curve with this method, the initial guess for the excited state calculations needs to be changed manually in a non-straightforward way. However, when using DO-GMF, the calculations can make use of the fact that the solutions giving the correct energy curve passing through the critical point conserve the saddle point order of 2. Since there is only one solution before the critical point, the saddle point order to target after the critical point is known without the user needing prior knowledge about the system. If a 2\textsuperscript{nd}-order saddle point is targeted with DO-GMF, the symmetry-broken energy curve is obtained systematically.

The greatest challenge of DO-GMF calculations of excited states is estimating the saddle point order of the target solution. This target saddle point order can be estimated crudely by using the number of negative elements of the diagonal Hessian approximation (eq.\ref{eq:diagonal_hessian}). This approximation typically underestimates the true target saddle point order, the more so, the larger the difference in dipole moments between the ground and target excited state, i.e.~the larger the charge transfer distance.\cite{Selenius2023} At the same time, the number of negative eigenvalues of the exact electronic Hessian at the excited state initial guess usually overestimates it. The estimate of the target saddle point order can be improved significantly by using a so-called freeze-and-release method based on constrained optimization.\cite{Selenius2023} The concept of this method is to freeze all degrees of freedom corresponding to the single-particle states of the ground state from and to which the electron is promoted to form the excited state initial guess. In doing so, all degrees of freedom corresponding to negative curvature at the target saddle point are frozen, meaning that a simple minimization can be carried out in the unconstrained subspace. This constrained minimization step functions as a pre-optimization which describes the effect of the excitation on those orbitals that are not directly excited. Afterward, the constraints are released, and a saddle point search is carried out in the full space starting from the constrained solution. Since this saddle point search is started from a much improved initial guess compared to simple excitation from the ground state, the robustness of the overall procedure is increased. It also improves the quality of the approximate Hessian, which is reevaluated at the constrained solution, both as a preconditioner for the optimization method and as a means of estimating the target saddle point order specifically for DO-GMF computations.
\section{Excited electronic states with self-interaction correction}\label{sec:ex-sic}
As demonstrated in section~\ref{sec:stat} for the H$_{2}^{-}$ ion, SIC introduces many more stationary points in the oo subspace of $\boldsymbol{\kappa}$ than just a minimum corresponding to a minimal SIE (maximal SIC). The SIE generally tends to be minimized if the occupied orbital space is maximally localized. Additional stationary points are introduced involving the oo orbital rotation space of the H$_{2}^{-}$ ion corresponding to the situation where the SIE is maximized (SIC minimized), which is the case when the oo subspace of $\boldsymbol{\kappa}$ is maximally delocalized. More realistic systems of scientific interest contain many more electronic degrees of freedom in both the oo and ov subspaces of $\boldsymbol{\kappa}$. In general, one can expect each orbital rotation degree of freedom in the oo subspace of $\boldsymbol{\kappa}$ to have at least one value that maximizes the SIE for that degree of freedom. As a result, the oo subspace contains a large amount of unphysical stationary points, all of which need to be avoided. The situation is additionally complicated by the fact that it is very difficult to diagnose or rule out whether such a saddle point in the oo subspace has been obtained from the orbital visualizations and single-particle energy values of a converged solution since the delocalization of the occupied orbitals is typically not humanly distinguishable from the fully localized case, especially if, and that is usually the case, the latter is not available for visual reference. Yet, such an analysis is crucial since solutions that fully minimize the SIE and equivalent solutions that maximize the SIE along some degrees of freedom usually show qualitatively similar, but quantitatively different energy and derivatives of the energy with respect to the positions of the nuclei, as becomes evident by comparing the energy curves of these pairs of solution in fig.~\ref{fig:h2-pes}.

SIC excited state optimizations can be performed with DO-MOM. Since DO-MOM can converge to any stationary point on the electronic energy surface and generally converges to whatever stationary point is closest to the initial guess, it is vital to provide an excited state initial guess that is as close to the global minimum in the oo subspace of $\boldsymbol{\kappa}$ as possible. Such an initial guess is created by optimizing the target excited state without SIC first, at which point the KS energy is invariant to the oo subspace of $\boldsymbol{\kappa}$. Afterward, a localization technique is applied to the solution without SIC. For this purpose, the Pipek-Mezey localization method\cite{Pipek1989} can be used. The excited state initial guess is then prepared by performing a minimization of the SIE in the oo subspace. This process may be repeated in regular step intervals during the DO-MOM optimization.

Even if this elaborate formalism is used, the DO-MOM optimization can still converge to a solution that does not fully minimize the SIE. To diagnose this problem computationally, the eigenvectors of the electronic Hessian corresponding to its negative eigenvalues at the converged solution can be evaluated, for instance with the generalized Davidson method. For a solution that minimizes the SIE none of these eigenvectors have significant contributions from the oo subspace of $\boldsymbol{\kappa}$. In practice, a threshold $t_{oo}$ on the percentage of each eigenvector that is allowed to be localized in the oo subspace can be defined according to
\begin{equation}\label{eq:oo-threshold}
    \left|\sum_{i}^{N_{oo}}\vect{v}^{T}\vect{u}_{i}^{oo}\vect{u}_{i}^{oo}\right| < t_{oo}\,,
\end{equation}
where $N_{oo}$ is the number of unit vectors $\vect{u}_{i}^{oo}$ of the oo subspace. A practical value is $t_{oo} = 0.5$\,. If $t_{oo}$ is exceeded, the obtained solution must be considered unreliable. Recovering a solution that fully minimizes the SIE from a solution that does not is a non-trivial endeavor since the calculation is stuck at an incorrect stationary point, which is generally far from any other stationary point and as such, not a good initial guess for converging to a different stationary point. Since the energy is known to be maximal at a stationary point along an eigenvector of the Hessian with negative curvature, and it is also known that the energy should be minimized along the eigenvector because it is localized in the oo subspace of $\boldsymbol{\kappa}$, a minimizing line search in the direction of the eigenvector can be carried out and the result used to restart the saddle point search with DO-MOM in hopes of finding a different solution that minimizes the SIE.

Alternatively, the framework of the DO-GMF method provides the means to prevent convergence to saddle points that are not minima in the oo subspace of $\boldsymbol{\kappa}$ in the first place. In this case, the idea is to apply the threshold in eq.~\ref{eq:oo-threshold} at every optimization step before evaluating the modified gradient (eq.~\ref{eq:gmf}). If any of the target eigenvectors of the Hessian obtained with the generalized Davidson method exceed the threshold and are deemed localized in the oo subspace, they are discarded and the next higher eigenvectors are evaluated. These eigenvectors are checked again, and the process is repeated until the target number of lowest eigenvectors of the Hessian localized in the ov subspace of $\boldsymbol{\kappa}$ are found. It is necessary to perform this check at every optimization step since areas on the electronic energy surface exist where the lowest eigenvectors of the Hessian can be localized in the oo subspace, as is the case in fig~\ref{fig:h2-pes} for a distance of 0.8\,\AA\ between the hydrogen atoms. This method ensures that the SIE is fully minimized at every optimization step and therefore, systematically prevents convergence to saddle points in the oo subspace. DO-GMF applying this formalism is much less dependent on the initial guess for the excited state saddle point search and even converges to a solution that minimizes the SIE when started from a solution that does not fully minimize the SIE after adding some numerical noise to the initial guess. A flowchart of this robust DO-GMF algorithm is given in Appendix B. \added[id=YLAS, comment={R1C7}]{Details of the DO-GMF and DO-MOM algorithms without the proposed extensions described in this article can, additionally, be found in refs.~\citenum{Schmerwitz2023} and \citenum{Levi2020b}, respectively.} \added[id=YLAS, comment={R1C2}]{The modifications to the algorithm do not impair the scaling of the method with the number of basis functions, which is the same as that of ground state DFT.}
\section{Conclusion}
%
\begin{figure*}[h!]
    \centering
    \includegraphics[width = \textwidth]{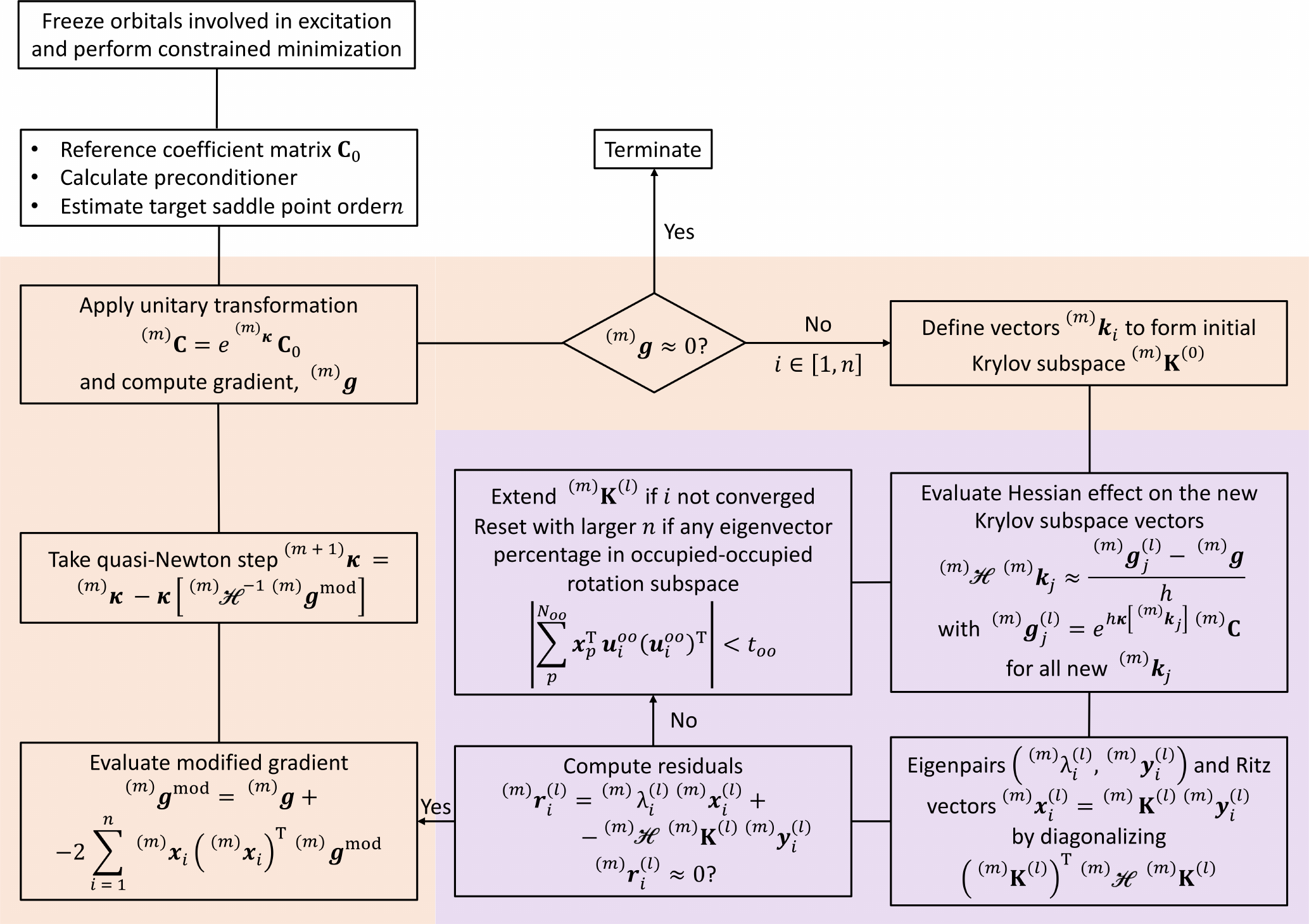}
    \caption{Flowchart of the DO-GMF algorithm. The method consists of an outer loop of direct optimization using quasi-Newton minimization (red) and an inner loop of determining the lowest $\vect{n}$ eigenvectors of the electronic Hessian with the generalized Davidson method (purple) to determine a modified gradient for the outer loop.}
    \Description{A flowchart is shown containing connected boxes with algorithmic steps of the direct optimization generalized mode following method. Steps of initialization are shown outside a main loop of optimization depicted with a light red background, while steps of a generalized Davidson method inner loop are shown with a light blue background. The reader is referred to Appendices A and B for a detailed description of the individual algorithmic steps.}
    \label{fig:flowchart}
\end{figure*}
Saddle point search algorithms for density functional calculations of excited electronic states with Perdew-Zunger self-interaction correction have been presented. The SCF method iteratively evaluates and diagonalizes the Hamiltonian matrix. This method has been designed for calculations of ground state minima and can suffer from variational collapse to lower-energy solutions if applied in excited state saddle point searches. To counteract this so-called variational collapse, a maximum overlap method can be used, but this method cannot always prevent collapse. The direct optimization method systematically converges to saddle points on the electronic energy surface and is, therefore, more robust than the SCF method together with the maximum overlap method. The latter is not strictly necessary for the direct optimization method, but typically employed as a fail-safe. While the standard direct optimization method can converge to any stationary point and usually converges to whatever stationary point is closest to the initial guess, the generalized mode following extension of direct optimization systematically targets saddle points of a specific order by following the $l$ lowest eigenvectors of the electronic Hessian up in energy. The saddle point order to target can be estimated accurately by performing a constrained minimization in the subspace of electronic degrees of freedom not involved in the excitation and evaluating the number of negative elements of a diagonal Hessian approximation. Typical exchange-and-correlation functionals employed in variational density functional calculation of excited electronic states do not fully compensate the self-interaction in the Coulomb interaction between the electrons, introducing a self-interaction error into the calculations. Perdew-Zunger self-interaction correction can be used to correct this error on a single-particle basis. While the uncorrected Kohn-Sham energy is invariant to the electronic degrees of freedom mixing the occupied orbitals, the correction lifts this invariance and introduces a large amount of stationary points localized in this space on the energy surface. Since only a minimum in this space truly minimizes the SIE, other stationary points in this space are unphysical and need to be avoided. The standard direct optimization method strongly depends on the initial guess, even if localization procedures are used. An extension of the generalized mode following method has been proposed that avoids following eigenvectors located in the occupied-occupied orbital rotation space and thereby, guarantees convergence to excited state solutions that minimize the self-interaction error.
\section{Appendix A: Generalized Davidson method}
The generalized Davidson method to calculate the lowest $n$ eigenvectors of the electronic Hessian has been presented in ref.~\citenum{Crouzeix1994}. The diagonal Hessian approximation (eq.~\ref{eq:diagonal_hessian}), $\matr{D}$, is evaluated during the initialization of the wave function optimization and available during the optimization. At the start of the generalized Davidson method, an initial Krylov subspace $\matr{K}$ is defined consisting of $n$ unit column vectors $\vect{k}_{i}$ along the elements of $\boldsymbol{\kappa}$ corresponding to the lowest $n$ elements of $\matr{D}$. $\matr{K}$ is orthonormalized with the modified Gram-Schmidt method after applying small numerical noise. A forward finite difference approximation is used to evaluate the effect of the electronic Hessian matrix, $\mathscrbf{H}$, on $\matr{K}$
\begin{equation}
\mathscrbf{H}\vect{k}_{j} \approx \frac{\nabla E\left(\matr{C}e^{h\boldsymbol{\kappa}\left[\vect{k}_{j}\right]}\right) - \nabla E\left(\matr{C}\right)}{h}\,
\end{equation}
with $\boldsymbol{\kappa}\left[\vect{k}_{j}\right]$ being the anti-Hermitian matrix containing the elements of the $j$\textsuperscript{th} vector of the Krylov subspace, $\vect{k}_{j}$, in its upper triangular part, $h$ the finite difference step size, and $\nabla E\left(\matr{C}\right)$ the energy gradient vector $\vect{g}$ calculated at the current LCAO coefficient matrix $\matr{C}$. A smaller representation of the eigenvalue problem is constructed and solved by forming and diagonalizing the Rayleigh matrix $\matr{K}^{\mathrm{T}}\mathscrbf{H}\matr{K}$. The full-dimensional approximation of the target eigenvectors, the Ritz vectors $\vect{x}_i = \matr{K}\vect{y}_i$, are evaluated from the obtained $n$ lowest eigenpairs ($\lambda_i$, $\vect{y}_i$). The residual vectors $\vect{r}_i=(\lambda_i\matr{I}-\mathscrbf{H})\vect{x}_i$ are formed and serve as the convergence criteria of the algorithm. Since they tend to zero, a maximum component of 0.01\,E\textsubscript{h} is applied as the convergence threshold for the residual of each target eigenvector. If the target eigenvector is not converged, the Krylov subspace is extended in the direction of the preconditioned residual vector
\begin{equation}
\matr{P}_i=(\lambda_i\matr{I}-\matr{D})^{-1}\,,
\end{equation}
where $\matr{I}$ is the identity matrix. As in ref.~\citenum{Sharada2015}, the elements of $\matr{P}$ are set to a threshold of $-0.1$\,E\textsubscript{h} if they exceed it to ensure that the preconditioner is negative-definite, so that the Davidson method converges to the lowest eigenpairs. Only the effect of the Hessian on the most recently added vectors in $\matr{K}$ needs to be evaluated at every iteration. The cost of the eigendecomposition of the Rayleigh matrix is kept negligible by resetting the Krylov subspace with the current approximate eigenvectors $\vect{x}_{i}$ and their preconditioned residual vectors if the dimensionality of $\matr{K}$ becomes too large. If convergence is signaled for all target eigenvectors, the modified gradient (eq.~\ref{eq:gmf}) is formed, and an optimization step is taken. Subsequent Davidson cycles make use of the eigenvectors found at the previous optimization step to form the initial Krylov subspace to accelerate convergence.
\section{Appendix B: Flowchart of the direct optimization generalized mode following algorithm}
The flowchart of the DO-GMF method for saddle point searches to determine excited state solutions in density functional calculations with SIC is shown in fig.~\ref{fig:flowchart}. The initial guess for the excited state saddle point search is generated from the ground state orbitals by promoting one or more electrons from an occupied single-particle state to a virtual one. The orbitals corresponding to these single-particle states are frozen, and a constrained minimization is performed in the remaining subspace. The reference matrix of LCAO coefficients, the diagonal approximation to the Hessian (eq.~\ref{eq:diagonal_hessian}), and the estimate of the target saddle point order of the excited state are reevaluated at the constrained solution, and the constraints are released. The DO-GMF method starts by applying the current unitary transformation to $\matr{C}_{0}$ (eq.~\ref{eq:unitary-transformation}) and computing the electronic gradient (eq.~\ref{eq:gradient}). Convergence is signaled if the norm of the gradient has sufficiently approached 0. If convergence has not been reached, the generalized Davidson method described in Appendix A is used to calculate the lowest $n$ eigenvectors of the electronic Hessian.
The eigenvectors are checked for their percentage of localization in the oo subspace of $\boldsymbol{\kappa}$. Any eigenvector that is localized in the oo subspace more than a threshold of $t_{oo}$ (eq.~\ref{eq:oo-threshold}) is discarded and the Davidson method continued targeting the next higher eigenvectors until $n$ eigenvectors localized in the ov subspace of $\boldsymbol{\kappa}$ are obtained. A value of $t_{oo} = 0.5$ is found to be sufficient in practical calculations. This threshold is applied to ensure that the optimization converges to a saddle point that minimizes the SIE. The modified gradient (eq.~\ref{eq:gmf}) is evaluated and a quasi-Newton step with the L-BFGS update formula is taken. 
\begin{acks}
This work was supported by the Icelandic Research Fund (grant agreements nos.\ 217751, 217734). The calculations were carried out at the Icelandic High Performance Computing Center (IHPC).
\end{acks}

\bibliographystyle{ACM-Reference-Format}
\bibliography{main}

\end{document}